# Is There A Real Estate Bubble in Switzerland? (diagnostic as of 2012-Q4)

Diego Ardila, Peter Cauwels, Dorsa Sanadgol, and Didier Sornette

Chair of Entrepreneurial Risks
Department of Management Technology and Economics
ETH Zurich

Scheuchzerstrasse 7
8092 Zurich
Switzerland

### Summary
We have analyzed the risks of possible development of bubbles in Switzerland's residential real estate market. The data employed in this work has been collected by comparis.ch, and carefully cleaned from duplicate records through a procedure based on supervised machine learning methods. The study uses the log periodic power law (LPPL) bubble model to analyze the development of asking prices of residential properties in all Swiss districts between 2005 and 2013. The results suggest that there are 11 critical districts that exhibit signatures of bubbles, and seven districts where bubbles have already burst. Despite these strong signatures, it is argued that, based on the current economic environment, a soft landing rather than a severe crash is expected.

## Introduction
The development of residential property prices over the past years in Switzerland has raised concerns about the existence of a bubble in this market. Key indicators, such as the ratio of home to rent prices, are deviating from the long-term equilibrium (UBS 2013a), whereas the direct exposure of banks to real estate has grown enough to pose a threat for the stability of the financial sector (SNB 2012). The situation is of great importance, as real estate volatility on large scale and intensity can have long-lasting and destructive effects for an economy. This was directly illustrated by the aftershocks of the burst of real estate bubbles in the U.S., Spain, and Ireland (Allen and Carletti 2010), and by the consequences of the bubble in Switzerland at the end of the 1980s. The Swiss real estate bubble, which was fueled by a decline in mortgage lending standards, caused a sharp drop in GDP of about 1.55 percent and resulted in severe price corrections, and widespread foreclosures (Bourassa, Hoesli, and Scognamiglio 2009). A repetition of this crisis today could have similar repercussions, as real estate assets represent 43.6 percent of the Swiss households' wealth according to data of the Swiss National Bank (SNB 2011).

It has been argued that the recent development of prices is due to a mismatch between supply and demand (UBS 2013b and Credit Suisse 2012). On the one hand, the demand has benefited from three factors: historically low interest rates; a sustained rate of immigration; and increasing real wages. On the other hand, the supply has had problems to keep up with the strong demand as it suffers from lengthy production times (Credit Suisse 2012).

Nevertheless, this conclusion is based on the same type of fundamental analysis that failed to detect the U.S. real estate bubble in 2007. At that time, it was boldly argued that there was little ground for bubble concerns as home prices had - allegedly - moved in line with increases in family income and declines in nominal mortgage interest rates (McCarthy and Peach 2004). Yet, the bubble burst and we are still indirectly bearing the consequences of it. This stresses the importance of a dynamical approach, and represents a strong case for prudence, since the diagnosis of bubbles remains a controversial topic with elusive targets.

In this context, we have started collaboration with comparis.ch in order to study the risks of a bubble in the residential Swiss real estate market. In this way, we received access to an exclusive set of data containing millions of records of asking prices giving us a unique view on the market, with a remarkable resolution in space. This document presents the analysis for each of the 166 districts of Switzerland (based on 2009 divisions) between 2005 and 2012.

## DATA PROCESSING

The data used in this analysis was collected by comparis.ch between January 2005 and December 2012. The property market division of comparis.ch gathers data from the 17 largest property portals in Switzerland, creating a rich view on the market, but also introducing a large and un-estimated number of duplicate ads (4'053'743 records present in the raw data). These duplicates advertise the same property, during the same period, and sometimes, with conflicting information. Within the scope of this study, the identification of the duplicates was crucial, as they could potentially affect the price indices.

We implemented a procedure based on the Support Vector Machine (SVM) algorithm (Scholkopf and Smola 2001) and string distance measures (Cohen, Ravikumar, and Fienberg 2003) in order to identify the duplicate ads. The procedure determined, in a given zip code and a given quarter, the ads that represented the same residential property by analyzing the similarity between their different attributes (e.g. their title, description, and number of rooms). In this study, we have only included ads with positive price and living space, as this information was essential to develop the price indices. In addition, ads with different prices were considered different since this study did not intend to track the price changes of the properties on sale.

## DEVELOPMENT OF THE PRICE INDEX AND SYNTHETIC VIEW

We have studied the development of prices in each of the 166 Swiss districts. In order to analyze the market, the ads in each district were categorized by type (i.e. apartment or house), and subsequently subdivided in three groups, according to their number of rooms, as described in table 1. The properties in each subgroup were aggregated quarterly using the median asking price and the median asking price per square meter for houses and apartments respectively.

| Property Type | House | | Apartments | |
|---|---|---|---|---|
| Measure | Median Asking Price | | Median Asking Price per Square Meter | |
| Size | Minimum # of Rooms | Maximum # of Rooms | Minimum # of Rooms | Maximum # of Rooms |
| Small | 1 | 4.5 | 1 | 3.5 |
| Medium | 5 | 6.5 | 4 | 5.5 |
| Large | ≥ 7 | | ≥ 6 | |

**Table 1: Classification of real estate ads.**

The application of the de-duplication procedure to the comparis.ch database classified approximately 550'000 houses and 460'000 apartments as having been for sale on the market between 2005Q1 and 2012Q4, which amount to a total of about 1 million residential properties. An overview of the market as of 1$^{st}$ of January 2013 is presented in figures 1 and 2. Cantonal median values are shown whenever not enough listings during the specified period were available for a district (less than 10 ads). The apartments in 70.5 percent of the districts exhibit a median asking price per square meter between 3'000 and 6'000 CHF. Entremont, Saanen and Maloja are the most expensive districts.

The corresponding disaggregation for medium size houses is shown in figure 2. Regardless of the size of the properties, the cantons of Genève, Zürich and Vaud are substantially pricier compared to the other cantons. In particular, 50 percent of the median asking prices for medium size houses over all the districts in these three cantons are greater than 1'200'000 CHF.

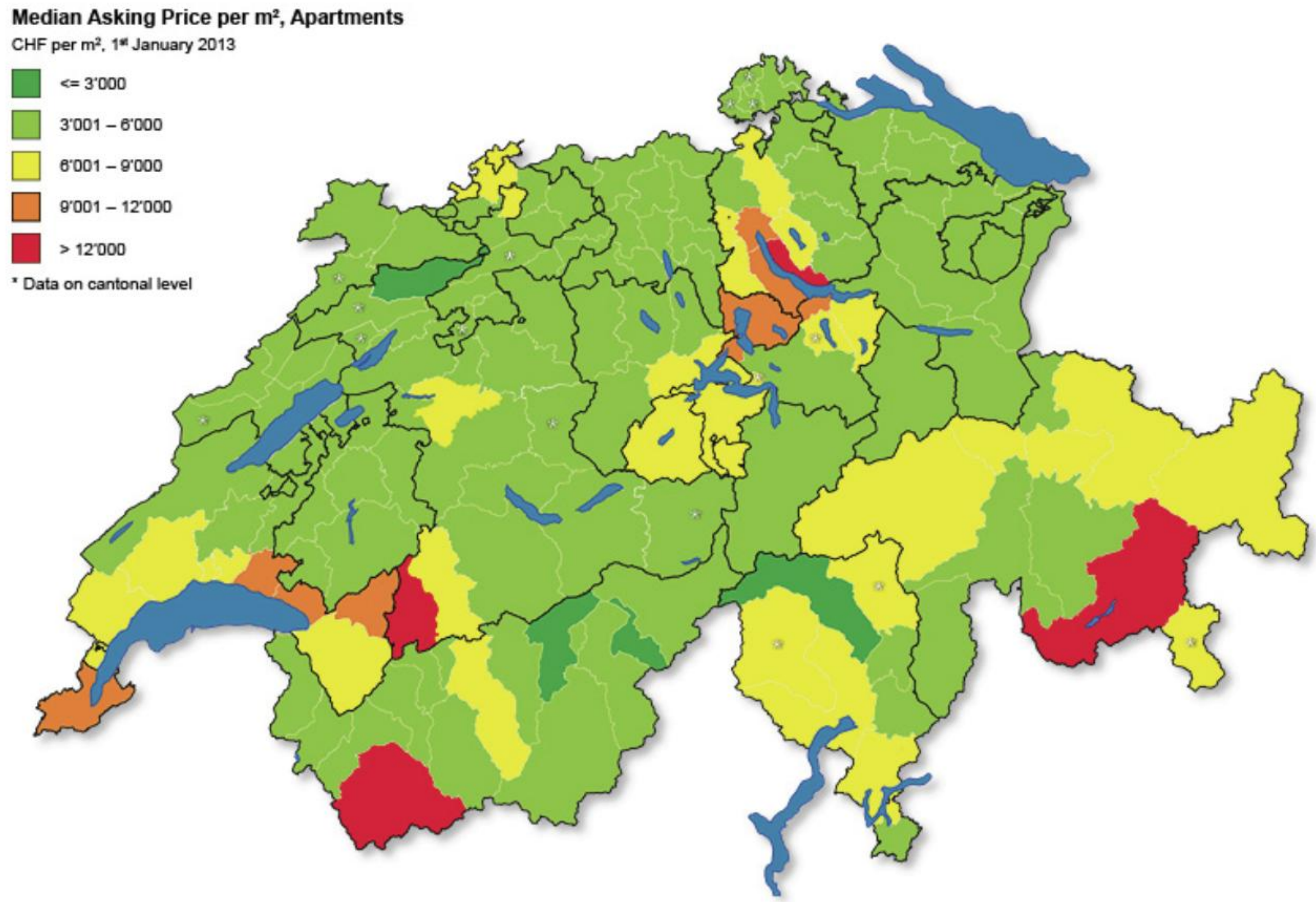

**Figure 1: Median Asking Price per Square Meter for Apartments in all Swiss Districts as of 1st January 2013.**

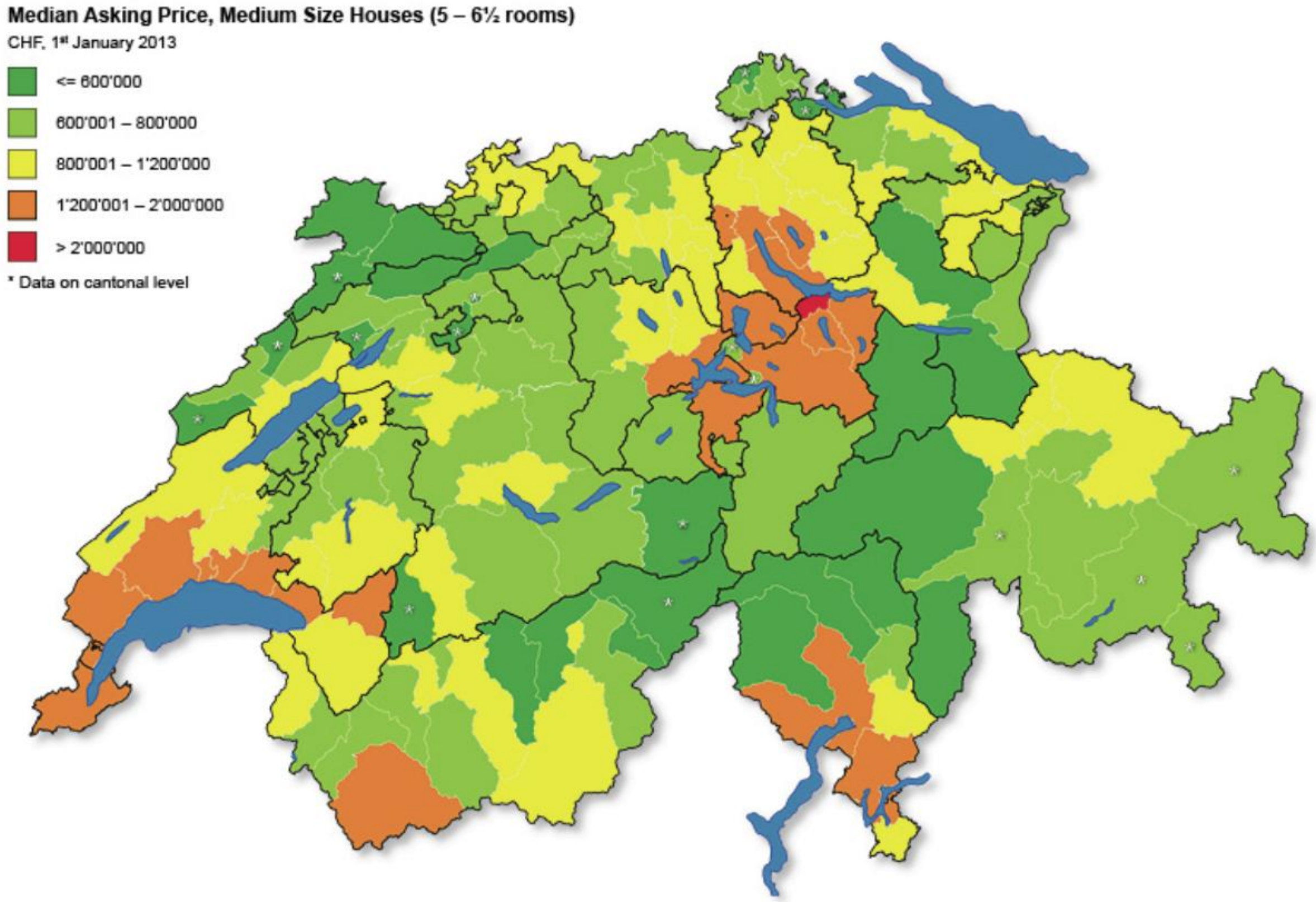

**Figure 2: Median Asking Price of Medium Size Houses in all Swiss Districts as of 1st January 2013.**

A heat map illustrating the price changes in apartments between 2007-Q1 and 2012-Q4 was developed to obtain a preliminary idea of the evolution of prices (figure 3). According to this data, 56 percent (93 out of 166) of the districts have undergone increases greater than 25 percent over this period. The most notable price change happened in Entremont, where the median asking price of apartments per square meter more than doubled. The districts of Genève and Zürich, along with the districts surrounding their lakes, as well as the touristic destinations in canton Graubünden all show a significant rise in asking price per square meter, mostly between 51 and 75 percent.

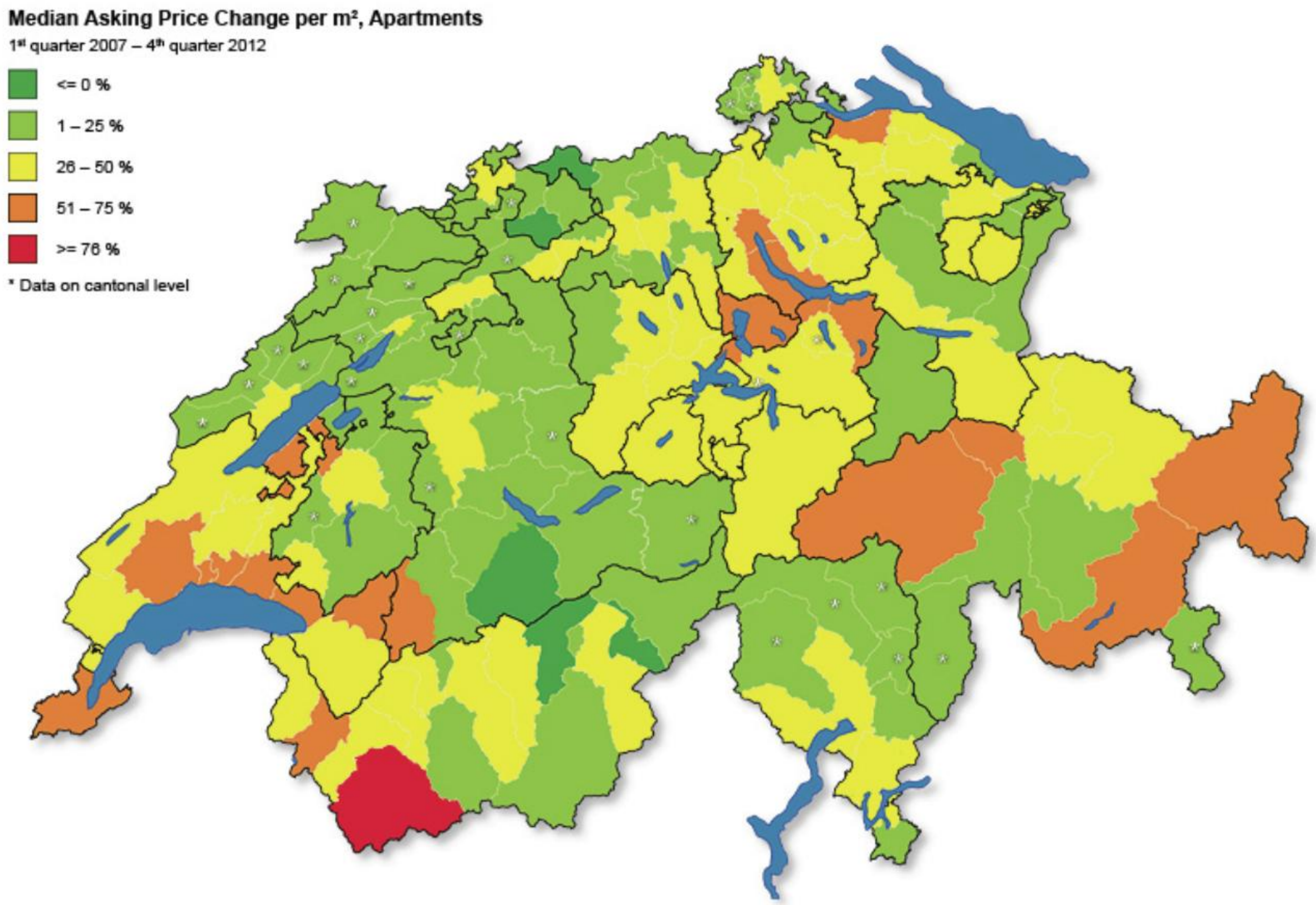

**Figure 3: Change In Median Asking Price Per Square Meter For Apartments In All Swiss Districts Between 2007 Q1 and 2012Q4.**

## BUBBLE DIAGNOSTIC

### Method: the log-periodic power law (LPPL) bubble model

The term "bubble" refers to a situation in which excessive public expectations of future price increases cause prices to be temporarily elevated (K. E. Case and Shiller 2003). Didier Sornette and Ryan Woodard (2010) illustrate the concept of housing price bubble as follows:

"During a housing price bubble, homebuyers think that a home that they would normally consider too expensive for them is now an acceptable purchase because they will be compensated by significant further price increases. They will not need to save as much as they otherwise might, because they expect the increased value of their home to do the saving for them. First-time homebuyers may also worry during a housing bubble that if they do not buy now, they will not be able to afford a home later."

We employed the log periodic power law (LPPL) bubble model to diagnose the risk of real estate bubbles in Switzerland. The LPPL model diagnoses a bubble as a transient, faster than exponential growth process, decorated with ever-increasing oscillations representing the low-frequency developing price volatility. In its microeconomic formulation, the model assumes a hierarchical organization of the market, comprised of two groups of agents: a group with rational expectations (the value investors), and a group of "noise" agents, who are boundedly rational and exhibit herding behavior (the trend followers). Herding is assumed to be self-reinforcing, corresponding to a nonlinear trend following behavior, which creates price-to-price positive feedback loops that yield an accelerated growth process. The tension and competition between the rational agents and the noise traders produces deviations around the growing prices that take the form of low-frequency oscillations, which increase in frequency due to the acceleration of the price and the nonlinear feedback mechanisms, as the time of the crash approaches.

In the LPPL model, a crash signals a change of regime, in which the prices stop rising, and take a different dynamics. This can be a swift correction, like a crash, but also a slow deflation or stagnation. In fact, a less violent and slower end of bubbles is a better representative characteristic of real estate markets since properties are durable goods that people tend to hold whenever falling prices are observed. In this case, the crash is more in the volume of transactions than in the price itself, which may take years to show a significant correction. Moreover, a crash is never a certain event but is characterized by a probability distribution. This is an essential ingredient for the bubble to exist, as it is only rational for financial agents to continue investing when the risk of the crash to happen is compensated by the positive return generated by the financial bubble, and when there exists a finite probability for the bubble to disappear smoothly. In other words, the bubble is only possible when the public opinion is not certain about its end.

Many examples of calibrations of financial bubbles with LPPLs have been reported in more than a decade since the LPPL model was introduced (Johansen, Sornette and Ledoit, 1999): see the list of articles at http://www.er.ethz.ch/publications/finance/bubbles_empirical. Jiang et al. (2010) in particular summarize well the theory and present several notable recent applications, in terms of the advance diagnostic of large bubbles that were later confirmed as they ended in momentous crashes. Among many others, the LPPL has been successfully used to diagnose in advance the U.S. real estate market bubble that burst in 2007, the oil bubble that crashed in 2008, and the Shanghai Composite index crashes in 2007 and 2009.

## CRITICAL REGIONS

We applied the LPPL methodology to all subcategories of properties (defined in table 1) as well as to the aggregated index for apartments over the period 2005Q1-2012Q4. In addition, as a back testing exercise, the model was fitted to the time series that only covered the period 2005Q1-2011Q4 in order to identify districts where regime changes have already occurred (i.e. bubbles that have already burst). We discarded calibrations that indicated a bubble end beyond the third quarter of 2014 (inclusive) not to search for the critical times of the end of bubbles too far into the future.

The result of this exercise is shown in figure 4. The districts labeled from 1 through 11 show signs of speculative bubbles with critical times between the first quarter of 2013 and the second quarter of 2014. As it was described above, the districts to watch correspond to regions where the model has predicted critical times in the year 2012, and the observed prices during the year suggest that this has actually happened. The bubbles in the latter have been diagnosed among all the apartments, except in the district of Dietikon where the bubble covered only small apartments.

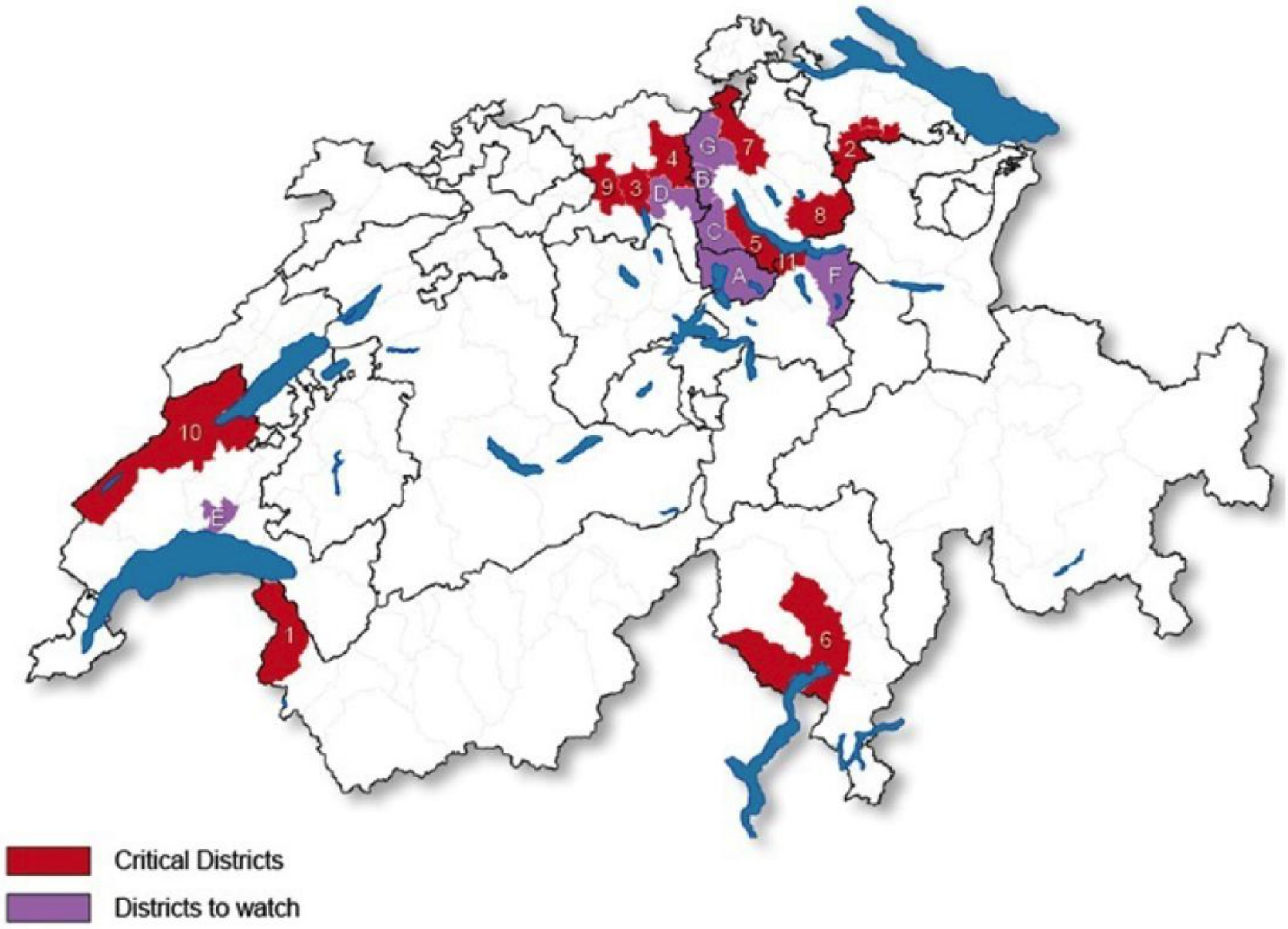

**Critical Districts**

| ID | District | Property Type | Size | Critical Time |
|---|---|---|---|---|
| 1 | Monthey | Apartments | Medium | 2013 Q1 – 2014 Q1 |
| 2 | Münchwilen | Apartments | Medium | 2013 Q1 – 2013 Q4 |
| 3 | Lenzburg | Houses | Medium | 2013 Q1 – 2014 Q1 |
| 4 | Baden | Apartments | All | 2013 Q1 – 2014 Q1 |
| 5 | Horgen | Apartments | All | 2013 Q1 – 2014 Q2 |
| 6 | Locarno | Apartments | All | 2013 Q1 – 2014 Q1 |
| 7 | Bülach | Apartments | Medium | 2013 Q1 – 2013 Q4 |
| 8 | Hinwil | Houses | Medium | 2013 Q1 – 2014 Q1 |
| 9 | Aarau | Houses | Medium | 2013 Q1 – 2014 Q1 |
| 10 | Jura-Nord vaudois | Houses | Medium | 2013 Q1 – 2014 Q1 |
| 11 | Höfe | Apartments | Medium | 2013 Q1 – 2014 Q1 |

**Districts to watch**

| A | B | C | D | E | F | G |
|---|---|---|---|---|---|---|
| Zug | Dietikon | Affoltern | Bremgarten | Lausanne | March | Dielsdorf |

**Figure 4: Critical districts and districts to watch.**

An example of the bubble analysis on the development of the median asking price per square meter for all apartments in Horgen (critical district) and the canton Zug (watch district) is shown in figure 5. Both regions exhibit the signals of bubbles according to the LPPL method: a super-exponential growth, accompanied by decorating oscillations. The gray area represents the 80% confidence interval of the critical time and the dotted lines represent possible LPPL scenarios. As can be seen in figure 5b, the change of regime in the canton Zug seems to have already occurred.

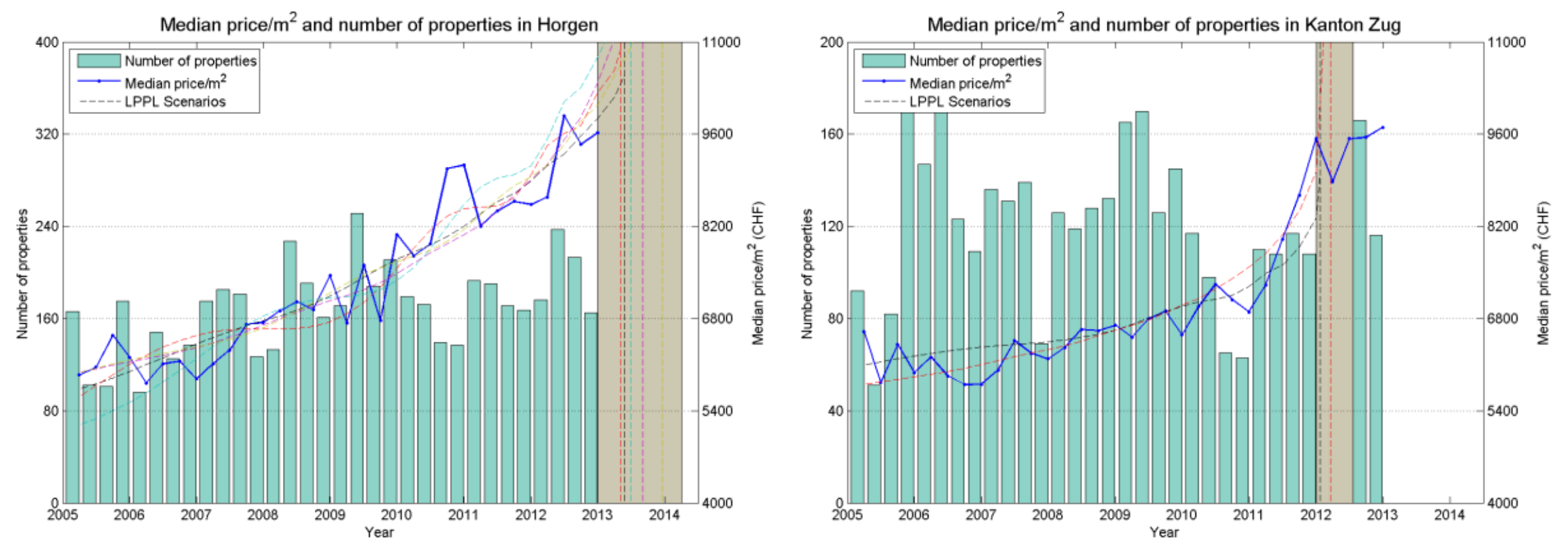

**(a)** **(b)**

**Figure5: (a) bubble in Horgen (critical district), all apartments; (b) bubble that burst in the canton Zug (watch district), all apartments.**

## COMPARISON TO OTHER STUDIES

We now compare our results with the UBS Bubble Index, which is published quarterly (UBS 2013a). This index comprises six different sub-indices that track the relationship between purchase and rental prices, the relationship between house prices and household income, the relationship between house prices and inflation, the relationship between mortgage debt and income, the relationship between construction and gross domestic product (GDP), and the proportion of credit applications by UBS clients for residential property not intended for owner occupancy. The selection of exposed regions is further conducted using a multi-level process that considers the size of the regional population and the property price data. In its 2012Q4 report, the index rose to 1.11 from 1.02, and highlighted seventeen exposed districts and nine monitored ones.

A one to one comparison with this index is not possible as the UBS index does not make any claim regarding the future development of prices, and even defines some regions differently. However, table 2 presents a simplified parallel, in which the districts depicted by the LPPL model and the UBS Bubble Index are directly contrasted. The LPPL model provides completely new information about nine districts (eight critical districts, and one to watch district), but does not report a critical situation in nineteen districts, where UBS points current exposure or need for monitoring.

| | UBS exposed districts | UBS monitored districts | Not reported by UBS |
|---|---|---|---|
| LPPL critical districts | Horgen(Zimmerberg) Höfe (March) | Bülach(partially in Glattal-Flurttal) | Monthey<br>Müchwillen<br>Lenzburg<br>Baden<br>Hinwil<br>Locarno<br>Aarau<br>Jura-Nord valudois |
| LPPL to watch districts | Zug<br>Dietikon(Limmatal)<br>March<br>Lausanne | Dielsdorf (partially in Glattal-Flurttal) | Bremgarten |
| Not reported by the LPPL model | Prättigau-Davos<br>Bernina (Oberengadin)<br>Maloja (Oberengadin)<br>Geneve<br>Nyon<br>Morges<br>Lavaux-Oron (Vevey)<br>Arlesheim (Unteres Basielbieg)<br>Saanen(Saanen-Obersimmental)<br>Zürich<br>Meilen(Pfannestiel) | Basel Stadt<br>Luzern<br>Appenzell-Innerhoden<br>Nidwalden<br>Uster<br>Gersau (Innerschvyz)<br>Kussnacht (Innerschvyz)<br>Schvyz (Innerschvyz) | |

**Table 2: Comparison between the LPPL results and the UBS Bubble Index (whenever the name of a UBS region does not coincide with a district, the original name is shown between parentheses. Districts are *partially monitored* or *partially exposed* if they are not fully covered by a UBS region).**

We looked closely at the exposed districts reported by the UBS Bubble Index that were not identified by the LPPL model, as this could imply an overlook of latent threats. Among these discrepancies, arguably, the most prominent is the absence of the cantons of Geneva, Vaud, and Graubünden, consistently reported by the UBS Bubble Index as risky zones. Our assessment found moderate or no bubble signatures in these regions. The reason for this can be found both in the data and in the methodology. First of all, the method applied by UBS compares the values of a region with those of all the country to determine the extent of a regional exposure. As a result, overpopulated regions or those that have historically exhibited above-average prices will tend to be consistently reported in the exposed category. In contrast, the LPPL model prioritizes the price-dynamics, requiring a faster than exponential growth to diagnose a price development as a bubble. In other words, our methodology emphasizes the information contained in the price dynamics and compares a district to itself at a previous time rather than to all its peers, rendering our analysis free of bias from the many factors that make different regions often hard to compare. The condition of a faster than exponential price growth is essential in our methodology, and is not fulfilled in the cantons of Geneva and Vaud, or in the district of Arlesheim. In these regions, the development of prices resembles a linear growth, which does not satisfy our definition of a bubble. For example, according to the comparis.ch database, the median asking price per square meter of apartments in the canton of Geneva has increased on average by 156 CHF per quarter during the observed period. Thus, these cantons do not show the typical signals of the bubbles identified by the LPPL model. A similar situation happened for the apartments of Zürich and Meilen.

The differences in Obersimmental and the two districts of Graubünden are bound to other reasons. The number of advertised properties that were available in the comparis.ch database for these districts over several quarters is very few or zero. Hence, it is not possible to draw any statistical conclusion about their prices, and the results of this study should not undermine the alarms raised by the UBS index. Having mentioned this, it is also worth noting that the low number of properties might not be an issue specific to our dataset, but rather a predominant characteristic of these locations, which serve mainly as luxurious or touristic destinations. If this was the case, the consequences of a bubble in these regions would be marginal for the overall economy as very few people would be affected.

The exception in Graubünden is the apartments in the district of Davos, which have seen a very liquid market. There, the situation needs further monitoring as the prices grew sharply during two years, starting from the beginning of 2010, and then stabilized. Although there were notable price increases, the acceleration occurred only during a very short period and therefore the region does not exhibit a signature of a bubble according to our model.

## CONCLUSION

Despite the observation of strong bubbles signals in several districts, there is no reason to panic. The current economic environment suggests that a soft landing of prices instead of a sharp correction can be expected. The main reasons for this claim are the following.

First, the rising property prices in Switzerland have not been accompanied by a boom in the construction sector, as was the case for the bubbles in the U.S., Ireland and Spain. In these countries, high supply introduced a construction boom that contributed to a stronger correction of high prices (Allen and Carletti 2010). On the contrary, the construction sector in Switzerland keeps moving slowly, and stays below historic averages. The vacancy rate in turn has stagnated at a low level, presenting a marginal increase of only one percent during the last year: from 38'420 empty apartments in June 2011 to 38'920 empty apartments in June 2012 (FSO 2012).

Second, the SNB is already issuing early and urgent measures to control the market. In February 2013, it ordered banks to hold a countercyclical capital buffer amounting to one percent of their risk weighted assets, backed by residential properties in Switzerland (SNB 2013). With this policy, the central bank is directly aiming to reduce the exposure of banks to real estate, which has proved a key amplifier of previous crashes (Hilbers, Lei, and Zacho 2001). It is estimated that the new policy will impact as much as 25 percent of the country's total mortgage volume (Bloomberg 2013), affecting especially Raiffeisen and regional banks, as most of their assets are mortgages.

Third, unlike the burst of the real-estate market bubble in Switzerland during the 1980s, which was fueled by a decline in mortgage lending standards (Westernhagen et al. 2004), Swiss banks are seeking to implement more conservative practices. The Swiss Financial Market Supervisory Authority FINMA approved a new set of minimum requirements for mortgage financing, drawn up by the Swiss Bankers Association (SBA). The new regime, which came into effect on July 2012, for the first time requires a minimum 10 percent down payment from the own borrower's funds without using the 2nd pillar of their retirement fund when purchasing a property and demands mortgages to be paid down to two thirds of the lending value within 20 years (SBA 2012). This new scheme should prevent households from taking greater risks, as they will be unable to overuse the money from their pension funds to make the down payment and will be pressed to reduce the burden of the debt.

In light of this reasoning, a severe crash in the identified critical and to be watched districts is less likely, and a soft landing or stagnation of prices is a more probable scenario. Yet, as the vigorous demand in 2013 is an economic reality, the possible change of regime in these districts might arguably be accompanied by increasing price pressures in their adjacent districts. This is plausible, not only because contagious effects have been observed in other housing market bubbles (Roehner 1999, Fry 2009), but also because immigrants, which represent an important driver of the current demand, are traditionally more flexible and willing to travel farther distances when looking for a place to live.

Having said this, it is also important to keep in mind that the impact of the preventive measures is yet to be seen. Not only there is no consensus concerning the role that central banks should play during bubble regimes (see Roubini 2006 and Posen 2006 for the main arguments), but also there is uncertainty regarding the strength and appropriate calibration of these measures (Central Banking Newsdesk 2013). The fact that the monetary policy of the Swiss National Bank is anchored to the international and in particular the European milieu only makes it harder to exert contra-cyclical pressure on the real estate market by interest rate policy. Indeed, interest rate will arguably remain low for an extended period of time due to the general indebtedness of European countries and the stagnant evolution of their economies (Sornette and Cauwels, 2012).

Moreover, the overall economic situation remains challenging and an exogenous shock cannot be discarded. Nonetheless, the results of this study extend only to endogenous crashes (Sornette et al. 2011). Thus, possible shocks such as the adverse scenario contemplated by the Financial Stability Report of the SNB (SNB 2012), which includes a sharp escalation of the European debt crisis that could lead to a deep recession in Switzerland, are beyond the scope of this analysis.

## Acknowledgement

The authors would like to thank comparis.ch and the Commission for Technology and Innovation (CTI) for their support for this project.

## References

Allen, Franklin / Carletti, Elena (2010): «An Overview of the Crisis: Causes, Consequences, and Solutions». In: International Review of Finance 10.1, pp. 1–26.

Bloomberg: «Swiss Property Bubble Concern Seen Prompting Tightening». Url: http://www.bloomberg.com/news/2013-02-13/swiss-property-bubble-concern-seen-prompting-tightening.html (visited on 02/22/2013).

Bourassa, Steven / Hoesli, M.E.R. / Scognamiglio, Donato (2009): «Housing finance, prices, and tenure in Switzerland». In: Swiss Finance Institute Research Paper, pp. 09-16.

Case, Karl E / Shiller, Robert J (2003): «Is there a bubble in the housing market?». In: Brookings Papers on Economic Activity 2003.2, pp. 299–362.

Central Banking Newsdesk (2013): «Swiss board member supports counter-cyclical capital buffer». Url : http://www.centralbanking.com/central-banking/speech/2203857/swiss-board-member-supportscountercyclical-capital-buffer (visited on 03/01/2013).

Cohen, William W / Ravikumar, Pradeep / Fienberg, Stephen E (2003): «A comparison of string distance metrics for name-matching tasks». In: Proceedings of the IJCAI-2003 Workshop on Information Integration on the Web (IIWeb-03), pp. 73-78.

Credit Suisse (2012): «Real Estate Monitor Q4-2012». Url : http://www.s-schofield.com/Files/64e10d77d0721b2fff13b4ab2068149b.pdf (visited on 02/22/2013).

Fry, John M (2009): «Bubbles and contagion in English house prices». Url: http://mpra.ub.uni-muenchen.de/17687/1/Fryhousing.pdf (visited on 02/22/2013).

FSO (2012): «Leerwohnungsziffer stagniert auf tiefem Niveau». Url: http://www.bfs.admin.ch/bfs/portal/de/index/themen/09/22/press.html?pressID=8265 (visited on 02/22/2013).

Hilbers, Paul Louis Ceriel / Lei, Qin / Zacho, Lisbeth (2001): «Real Estate Market Developments and Financial Sector Soundness». In: Vol. 1. International Monetary Fund.

Jiang, Zhi-Qiang / Zhou, Wei-Xing / Sornette, Didier / Woodard, Ryan / Bastiaensen, Ken / Cauwels, Peter (2010): «Bubble diagnosis and prediction of the 2005–2007 and 2008–2009 Chinese stock market bubbles». In: Journal of economic behavior & organization 74.3, pp. 149–162.

Johansen, Anders, Didier Sornette and Olivier Ledoit (1999): «Predicting Financial Crashes using discrete scale invariance». In: Journal of Risk, vol. 1, number 4, 5-32 (1999).

McCarthy, Jonathan / Peach, Richard (2004): «Are home prices the next bubble?». In: Economic Policy Review 10.3.

Posen, Adam S (2006). «Why Central Banks Should Not Burst Bubbles». In: International Finance 9.1,pp. 109–124.

Roehner, B.M. (1999): «Spatial analysis of real estate price bubbles: Paris, 1984–1993». In: Regional science and urban economics 29.1, pp. 73–88.

Roubini, Nouriel (2006). «Why Central Banks Should Burst Bubbles». In: International Finance 9.1, pp. 87–107.

Scholkopf, Bernhard / Smola, Alexander J. (2001): «Learning with Kernels: Support Vector Machines, Regularization, Optimization, and Beyond». In: MIT Press, Cambridge, MA, USA. ISBN: 0262194759.

SNB (2011): «Household Wealth 2011» Url: http://www.snb.ch/ext/stats/wph/pdf/en/Verm_priv_Haush.pdf (visited on 02/22/2013).

SNB (2012): «Swiss National Bank Financial Stability Report 2012». Url: http://www.snb.ch/en/mmr/reference/stabrep_2012/source/stabrep_2012.en.pdf (visited on 02/22/2013).

SNB (2013): «Countercyclical capital buffer: proposal of the Swiss National Bank and decision of the Federal Council» Url: http://www.snb.ch/en/mmr/reference/pre_20130213/source/pre_20130213.en.pdf (visited on 02/22/2013).

Sornette, Didier and Peter Cauwels (2012): «The Illusion of the Perpetual Money Machine». In: Notenstein Academy White Paper Series (http://ssrn.com/abstract=2191509).

Sornette, Didier / Woodard, Ryan (2010): «Financial bubbles, real estate bubbles, derivative bubbles, and the financial and economic crisis». In: The Proceedings of APFA7 (Applications of Physics in Financial Analysis), Econophysics Approaches to Large-Scale Business Data and Financial Crisis, pp. 101–148.

Sornette, Didier / Woodard, Ryan / Yan, Wanfeng / Zhou, Wei-Xing (2011): «Clarifications to Questions and Criticisms on the Johansen-Ledoit-Sornette Bubble Model». In: Swiss Finance Institute Research Paper, pp. 11-29. (http://ssrn.com/abstract=2191524).

Swiss Bankers Association (2012): «Richtlinien betreffend Mindestanforderungen bei Hypothekarfinanzierungen». Url: http://shop.sba.ch/999985_d.pdf (visited on 02/22/2013).

UBS (2013a): «Swiss Real Estate Bubble Index: 4th quarter 2012». Url: http://www.ubs.com/global/en/wealth_management/wealth_management_research/bubble_index.html (visited on 02/22/2013).

UBS (2013b): «UBS Outlook Switzerland». Url: http://www.ubs.com/global/en/wealth_management/wealth_management_research/ubs_outlook_ch.html (visited on 02/22/2013).

Westernhagen, Natalja / Harada, Eiji / Nagata, T / Vale, B and others (2004): «Bank failures in mature economies». In: Basel Committee on Banking Supervision Working Paper 13.